\documentclass[aps,preprint]{revtex4}%
\usepackage{amsfonts}
\usepackage{amsmath}
\usepackage{amssymb}
\usepackage{graphicx}%
\begin{document}
\preprint{HEP/123-qed}
\title{Two-bands effect on the superconducting fluctuating diamagnetism in MgB$_{2}$}
\author{L.Roman\`{o}}
\affiliation{Department of Physics and Unit\`{a} INFM --CNR, University of Parma, Parco
Area delle Scienze 7A, 43100 Parma (Italy)}
\author{A.Lascialfari, A.Rigamonti}
\affiliation{Department of Physics and Unit\`{a} INFM --CNR, University of Pavia, Via Bassi
n.6 ,I-27100, Pavia (Italy)}
\author{I.Zucca}
\affiliation{Department of Physics and Unit\`{a} INFM --CNR, University of Pavia, Via Bassi
n.6 ,I-27100, Pavia (Italy)}
\affiliation{}

\begin{abstract}
The field dependence of the magnetization above the transition temperature
$T_{c}$ in MgB$_{2}$ is shown to evidence a diamagnetic contribution
consistent with superconducting fluctuations reflecting both the $\sigma$ and
$\pi$ bands. In particular, the upturn field $H_{up}$ in the magnetization
curve, related to the incipient effect of the magnetic field in quenching the
fluctuating pairs, displays a double structure, in correspondence to two
correlation lengths. The experimental findings are satisfactorily described by
the extension to the diamagnetism of a recent theory for paraconductivity, in
the framework of a zero-dimensional model for the fluctuating superconducting
droplets above $T_{c}.$


\end{abstract}
\maketitle

\clearpage On approaching the transition temperature $T_{c}$ from above,
superconducting fluctuations (SF) occur and, while the average order parameter
is zero, one has $\left\langle \left\vert \psi\right\vert ^{2}\right\rangle
^{1/2}\neq0$. Therefore local pairs are generated, lacking of long-range
coherence and decaying with a characteristic time which increases for
$T\rightarrow T_{c}^{+}$. An interesting way to detect the SF is by means of
the related magnetic screening, through a diamagnetic contribution to the
magnetization. In its simplest form this contribution is written
$-M_{dia}(H,T)\approx\frac{e^{2}}{mc^{2}}n_{c}\xi^{2}(T)H$, $n_{c}=\left\vert
\psi\right\vert ^{2}$ being the number density of superconducting pairs and
$\xi(T)$ the coherence length. On cooling towards $T_{c}^{+}$, because of the
divergence of $\xi(T)$ $|M_{dia}|$ can be expected to enhance. On the other
hand strong magnetic fields, comparable to the critical field $H_{c2}$, must
evidently suppress the SF. Thus the isothermal magnetization curves
$-M_{dia}(H,T=const)$ exhibit an upturn in the field dependence. According to
a model of fluctuating superconductive particles of size $d<<\xi(T)$ and
outside the critical region, the upturn field $H_{up}$ is about inversely
proportional to the square of $\xi(T)$ \cite{Tinkham,Larkin}.

In conventional metallic BCS superconductors the diamagnetism above $T_{c}$
and the effect of the field in quenching the fluctuating pairs were detected a
long ago, by means of magnetization measurements as a function of temperature
at constant fields \cite{gollub}. More recently, successful detection of the
detailed field dependence of $M_{dia}$ has been obtained in the new
superconductor MgB$_{2}$ \cite{Alemgb2}.

\qquad In Fig.3 of the paper by \textit{Lascialfari et al.} \cite{Alemgb2} an
anomaly was noticeable: the curve of the reduced magnetization at
$T\,=\,39.5\,K$ (while $T_{c}\simeq39.05\,K$) showed an unexplained double
structure. Later on, in view of the established two-bands character of the
superconductivity in MgB$_{2}$ \cite{Kortus}, it was suspected that the double
structure in the magnetization curve could be related to the two
superconducting bands in this compound. The motivation of the present letter
is mainly related to this hypothesis. In the following we are going to show
that the fluctuating diamagnetism (FD) above $T_{c}$ does reflect the presence
of both the $\sigma$ and $\pi$ bands which from a variety of effects are known
to occur in MgB$_{2}$ below $T_{c}$.

New magnetization measurements, with high temperature and field resolution
have been carried out above $T_{c}=39.1\,K\pm0.04\,K$, on a high-purity sample
prepared by \textit{Palenzona et al.} (University of Genova). The diamagnetic
contribution was obtained by subtracting from the raw data the paramagnetic
contribution measured at $40K$ (where the SF are practically ineffective). The
data obtained in a powdered sample at three representative temperatures are
reported in Fig.1.

The field dependence of FD can be understood in the framework of
Ginzburg-Landau (GL) theory in a simple way, by resorting to a description
based on fluctuation-induced superconducting droplets of spherical shape, with
diameter of the order of the coherence length. For these droplets the so
called zero-dimensional (0-D) approximation can be used, by assuming an order
parameter no longer spatial dependent. Then an exact solution for the GL
functional is found in closed form, valid above the critical region and for
all field $H<<H_{c2}$. This was basically the framework used in the discussion
of the data in Ref.\cite{Alemgb2}.

Recently, SF in the presence of two superconducting bands has been studied in
regards of the paraconductivity and of the specific heat in MgB$_{2}$
\cite{Varla}. In this work a breakdown of the GL theory is hypotized, due to
two different coherence lengths for the $\sigma$ and $\pi$ bands. In
particular, the difference is relevant in the z-direction, i.e. $\xi_{\sigma
z}\ll$ $\xi_{\pi z}$. An effective coherence length $\widetilde{\xi}_{z}(T)$
is introduced, with $\xi_{\sigma z}<<\widetilde{\xi}_{z}(T)<<\xi_{\pi z}$, in
a large temperature range where a generalized non-localized GL model can be
used \cite{Varla}.\newline

A system with weakly-coupled two bands labelled by $1$ (in MgB$_{2}$ the
$\sigma$ band) and $2$ (in MgB$_{2}$ the $\pi$ band) is considered and pairs
can be formed only by electrons belonging to the same band. The pairing
correlation matrix contains non-diagonal terms, allowing the pair to transfer
from one band to the other. Dealing with the impurity scattering, it is known
that in MgB$_{2}$ the interband contribution is very weak, because of the
different parity of the bands. Thus the transfer process can be neglected and
only the scattering frequency $\nu_{1}$ and $\nu_{2}$ in the two bands is
considered. The inverse of the effective coupling matrix is introduced%
\[
\widehat{W}=\left(
\begin{array}
[c]{cc}%
W_{11} & W_{12}\\
W_{21} & W_{22}%
\end{array}
\right)
\]

such that $W_{11}W_{22}-W_{12}W_{21}=0$.

The free energy above $T_{c}$ is determined by the fluctuation propagator
\cite{Varla}%

\begin{equation}
F=-k_{B}TV\int\frac{d^{3}q}{\left(  2\pi\right)  ^{3}}\ln\frac{A}{\det
H_{\alpha\beta}}. \label{2B1}%
\end{equation}

where $H_{\alpha\beta}(q)$ is the linearized GL Hamiltonian density\newline%

\begin{equation}
H_{\alpha\beta}=\left(
\begin{array}
[c]{cc}%
\nu_{1}(\xi_{1,a}^{2}q_{a}^{2}+W_{11}+\epsilon) & -\nu_{2}W_{12}\\
-\nu_{1}W_{21} & \nu_{2}\left[  W_{22}+\epsilon+f\left(  \xi_{2z}^{2}q_{z}%
^{2}\right)  \xi_{2x}^{2}q_{\parallel}^{2}+g\left(  \frac{2}{\pi^{2}}\xi
_{2z}^{2}q_{z}^{2}\right)  \right]
\end{array}
\right)  \label{2B2}%
\end{equation}

with $\varepsilon=\frac{T-T_{c}}{T_{c}}$. Here $g(x)\equiv\psi(1/2+x)-\psi(x)$
and $f(x)=\frac{2}{\pi^{2}}g^{\prime}(x)$, while $\xi_{1,a}^{2}q_{a}^{2}%
=\xi_{1x}^{2}q_{x}^{2}+\xi_{1y}^{2}q_{y}^{2}+\xi_{1z}^{2}q_{z}^{2}$.

At a temperature close to $T_{c}$, where $\widetilde{\xi}_{z}(T)>>\xi_{2z}%
(0)$, only the long wave-length fluctuations of the order parameter can be
taken into account. In this case, in the 0-D approximation and when written in
terms of the effective coherence length $\widetilde{\xi}(T)$, the
magnetization takes the form \cite{Larkin},%

\begin{equation}
M_{dia}(\varepsilon,H)=-K_{B}\,TH\frac{2\pi^{2}\widetilde{\xi}^{2}d^{2}%
/5\Phi_{0}^{2}}{\left(  \varepsilon+2\pi^{2}\widetilde{\xi}^{2}H^{2}%
d^{2}/5\Phi_{0}^{2}\right)  } \label{2B3}%
\end{equation}

Here $\Phi_{0}$ is the flux quantum and $d$ the size of fluctuating droplets,
that can be assumed of the order of $\widetilde{\xi}(T)$. Eq.\thinspace
(\ref{2B3}) predicts a single upturn in the field dependence of $M_{dia}$, at
a field $H_{up}\simeq\sqrt{5}\varepsilon\Phi_{0}$ /$\pi\widetilde{\xi}^{2}.$

For temperatures not too close to $T_{c}$, the short wave-length fluctuations
are accounted for by a large value of the argument in the $g$ and $f$
functions in Eq.\thinspace(\ref{2B2}). In order to obtain the fluctuating
magnetization, the full expression of the free energy has to be derived. By
assuming $\gamma=\dfrac{\xi_{1z}^{2}}{\xi_{1x}^{2}}$ as a measure of the
anisotropy between the plane and z-axis and neglecting the in-plane
anisotropy, one writes
\begin{align}
{\xi_{1,a}^{2}q_{a}^{2}}  &  {=\frac{H^{2}}{H_{c1x}^{2}}}\left(
{1+\frac{\gamma}{2}\sin^{2}\theta}\right)  {\nonumber}\label{2B4}\\
{\xi_{2x}^{2}q_{\parallel}^{2}}  &  {=\xi_{2x}^{2}}\left(  {q_{x}^{2}%
+q_{y}^{2}}\right)  {=\frac{H^{2}}{H_{c2x}^{2}}}\\
{\xi_{2z}^{2}q_{z}^{2}}  &  {=\frac{H^{2}}{H_{c2z}^{2}}\sin^{2}\theta
\nonumber}%
\end{align}
where $\theta$ is the angle between $H$ and the $z$-axis. In MgB$_{2}$ it can
be assumed \cite{Varla} $\xi_{1x}\simeq\xi_{2x}$ so that the critical fields
become $H_{c1x}^{2}=H_{c2x}^{2}=\dfrac{8\phi_{0}^{2}}{\pi^{2}\xi_{1x}^{2}%
d^{2}}$ and $H_{c2z}^{2}=\dfrac{16\phi_{0}^{2}}{\pi^{2}\xi_{2z}^{2}d^{2}}$.

From Eqs.(\thinspace\ref{2B1}\thinspace) and (\thinspace\ref{2B2}\thinspace) ,
by taking into account Eqs\thinspace(\ref{2B4}), a lengthly expression for the
angular-dependent diamagnetic magnetization is obtained. For external field
along the $z$-axis no double structure in the magnetization curves is present,
as it could be expected since only one effective correlation length is
involved:%
\begin{align}
M_{dia}(T  &  =const,\theta=0,H)=\nonumber\\
&  =-2k_{B}TVH\frac{\dfrac{1}{H_{cx}^{2}}\left(  1+\dfrac{\pi^{2}}{2}\right)
\epsilon+\dfrac{1}{H_{cx}^{2}}\left(  W_{22}+\dfrac{\pi^{2}}{2}W_{11}\right)
+\pi^{2}\dfrac{H^{2}}{H_{cx}^{4}}}{\left[  W_{11}+W_{22}+\dfrac{H^{2}}%
{H_{cx}^{2}}\left(  1+\dfrac{\pi^{2}}{2}\right)  \right]  \epsilon
+\dfrac{H^{2}}{H_{cx}^{2}}\left(  W_{22}+\dfrac{\pi^{2}}{2}W_{11}\right)
+\dfrac{H^{4}}{H_{cx}^{4}}}%
\end{align}

On the contrary for field in the $ab$-plane the magnetization turns out%
\begin{align}
M_{dia}(T  &  =const,\theta=\frac{\pi}{2},H)=\nonumber\\
&  =-\dfrac{\pi k_{B}Td^{3}H}{3S(H)}\Biggl[\left[  \left(  1+\frac{1}{2}%
\gamma\right)  +f\left(  \dfrac{\pi^{2}}{2}t\right)  +H^{2}f^{\prime}%
+H_{cx}^{2}g^{\prime}\right]  \frac{\epsilon}{H_{cx}^{2}}\\
&  +\dfrac{1}{H_{cx}^{2}}\left[  W_{11}+\dfrac{H^{2}}{H_{cx}^{2}}\left(
2+\gamma\right)  \right]  f\left(  \dfrac{\pi^{2}}{2}t\right)  +\dfrac{H^{2}%
}{H_{cx}^{2}}\left[  W_{11}+\dfrac{H^{2}}{H_{cx}^{2}}\left(  1+\frac{1}%
{2}\gamma\right)  \right]  f^{\prime}+\nonumber\\
&  \dfrac{1}{H_{cx}^{2}}\left(  1+\frac{1}{2}\gamma\right)  g\left(  t\right)
+\left[  W_{11}+\dfrac{H^{2}}{H_{cx}^{2}}\left(  1+\frac{1}{2}\gamma\right)
\right]  g^{\prime}+\dfrac{1}{H_{cx}^{2}}\left(  1+\frac{1}{2}\gamma\right)
W_{22}\Biggr]\nonumber
\end{align}

where $t=\frac{2}{\pi^{2}}\dfrac{H^{2}}{H_{c2z}^{2}}$ and $g^{\prime}$and
$f^{\prime}$ are the derivative of $g$ and $f$ \ functions with respect to
$H^{2}.$ In Eq.(6) the function $S(H)$ takes the form%

\begin{align*}
S(H)  &  =W_{11}+W_{22}+\dfrac{H^{2}}{H_{cx}^{2}}\left(  1+\frac{1}{2}%
\gamma\right)  +f\left(  \dfrac{\pi^{2}}{2}t\right)  \dfrac{H^{2}}{H_{cx}^{2}%
}+g\left(  t\right)  \epsilon+\\
&  +\dfrac{H^{2}}{H_{cx}^{2}}\left[  W_{11}+\dfrac{H^{2}}{H_{cx}^{2}}\left(
1+\frac{1}{2}\gamma\right)  f\left(  \dfrac{\pi^{2}}{2}t\right)  \right]
+\left[  W_{11}+\dfrac{H^{2}}{H_{cx}^{2}}\left(  1+\frac{1}{2}\gamma\right)
\right]  g\left(  t\right)  +\dfrac{H^{2}}{H_{cx}^{2}}\left(  1+\frac{1}%
{2}\gamma\right)  W_{22}%
\end{align*}

In Fig.2 the theoretical magnetization resulting from this equation is
reported for a representative value of the reduced temperature $\varepsilon$
and compared with the one derived for single band, for the effective
correlation length. It is noticeable how the shape of the magnetization curve
is affected by the two bands, displaying a double structure which resembles
two upturn fields.

Single crystals of MgB$_{2}$ large enough to yield a size of diamagnetic
signal above $T_{c}$ are not available and therefore our measurements have
been carried out in powders. Thus the data reported in Fig.1 can be considered
approximately correspondent to the theoretical case of $\theta\simeq\frac{\pi
}{2}$. From the inset in Fig.1 it is noted that for $T$ close to $T_{c}$ one
has a magnetization curve similar to the dashed one in Fig.2. This type of
field dependence, with a single upturn field, can be considered a signature of
the validity of the GL regime, for $\varepsilon\simeq3\times10^{-3}$.
Eq.\thinspace(\ref{2B3}) gives a satisfactory fitting of the data in
correspondence to a value $\widetilde{\xi}=22nm$, compatible with the
coherence length (see Ref.\cite{serventi}) .

For $T$ not too close to $T_{c}$, $M_{dia}$ should be discussed in terms of
the angular dependent full expression (not reported here), by averaging over
$\theta$. By using for \ a qualitative illustration the form of the
magnetization pertaining to $\theta=\frac{\pi}{2}$ a reasonable fit of the
data is obtained (Fig.3). In the fit we have used $\gamma\simeq0.02$. The
critical fields turn out $H_{c1x}=9.8\times10^{4}Oe$ and $H_{c2z}%
=11.5\times10^{2}Oe$, dependent on temperature through $d=$ $\widetilde{\xi
}(T)=\widetilde{\xi}(0)\varepsilon^{-1/2}$. $W_{11}$ and $W_{22}$ depend on
temperature. Far from $T_{c}$ the superconductive coupling in the $\pi$ band
is weak and this corresponds to a large $W_{22}$. In the inset of Fig.3 the
ratio $W_{11}/W_{22}$ is plotted as a function of the reduced temperature,
showing a dependence of the form $W_{11}/W_{22}\sim\varepsilon^{2}$. The ratio
of the magnetization values at the upturn fields $M_{up1}/M_{up2}$ shows the
same temperature dependence. This evidences that the SF involving the $\pi$
band are supressed at a field smaller than the ones related to the $\sigma$
band. By increasing the temperature the decrease of the coupling in the $\pi$
band produces a reduced contribution to the fluctuating magnetization
.\newline

By summarizing, in this work we have shown that the presence of the two SC
$\sigma$ and $\pi$ bands in MgB$_{2}$ has a noticeable effect also in the
fluctuating diamagnetism above $T_{c}$. By adapting to the fluctuating
diamagnetism the approach recently developed for the paraconductivity, a model
extending the Ginzburg-Landau formalism to include the effect of
non-evanescent magnetic field has been developed. In particular, by relying on
the zero-dimensional approximation for the superconducting droplets the model
has been proved to account for the basic aspects of the experimental findings.

\bigskip

\textbf{Acknowledgement}

\bigskip We are very grateful to Prof. A.A.Varlamov for stimulating
discussions and for its useful support in the calculations.

\bigskip

\textbf{Captions for figures}

Fig.1 Diamagnetic contribution $M_{dia}$ to the magnetization in MgB$_{2}$ as
a function of the magnetic field $H$ (after zero field cooling), for three
temperatures above $T_{c}=39.1\,\pm0.04\,K$. No difference was observed for
the magnetization in field-cooled condition at the same temperature (data not
reported). In the inset the data for $M_{dia}$ at a temperature close to
$T_{c}$ are reported, showing that for $H$ smaller than the upturn field
$H_{up\text{ }}$the GL law in finite field, namely $M_{dia}(H,T\approx
T_{c})\propto-\sqrt{H}$ (solid line), is well obeyed.

Fig.2 Comparison of the field dependence of the fluctuating magnetization
above $T_{c}$, for reduced temperature $\epsilon=8\times10^{-3}$, in the case
of a single superconducting band (dashed lines) and for the $\sigma$ and $\pi$
bands (solid line).

Fig.3 Fits of the magnetization curves at two representative temperatures
above $T_{c}$ in powered MgB$_{2}$ (see Fig.1) on the basis of the theoretical
expression (\thinspace Eq.\thinspace(6)\thinspace) in the text, according to
the generalized, non-local GL functional. In the inset the temperature
dependence of the ratios $\frac{W_{11}}{W_{22}}$ and $\frac{M_{dia}(H_{up}%
^{1},T)}{M_{dia}(H_{up}^{2},T)}$ are reported.The departure between the
theoretical prediction and the experimental data for $H\rightarrow0$ is likely
to be due to the subtraction procedure. In fact a very small amount of
ferromagnetic impurities could enhance the paramagnetic term for small fields.

\-\-
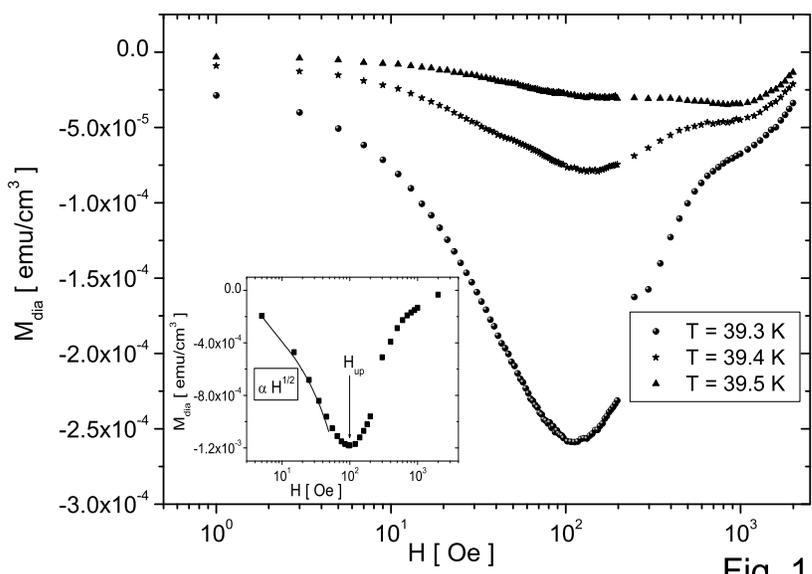
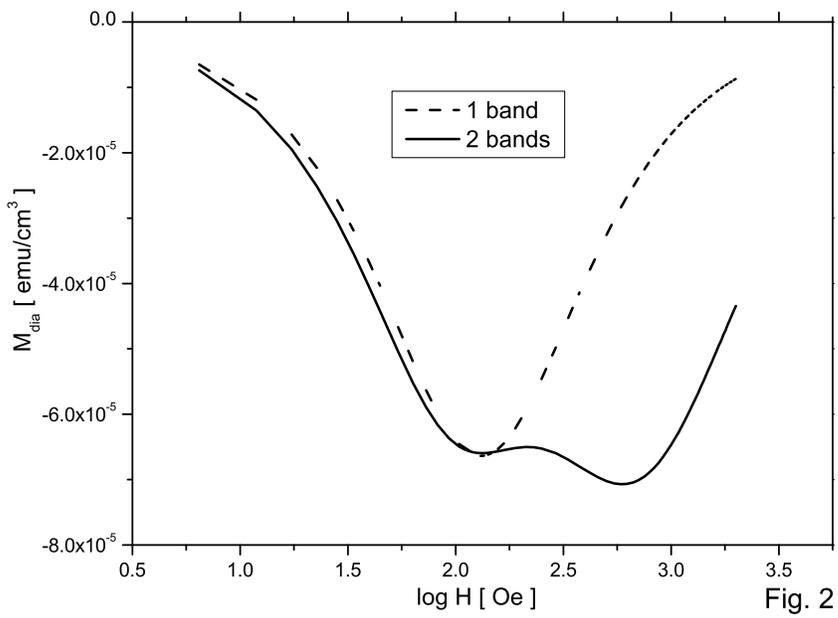
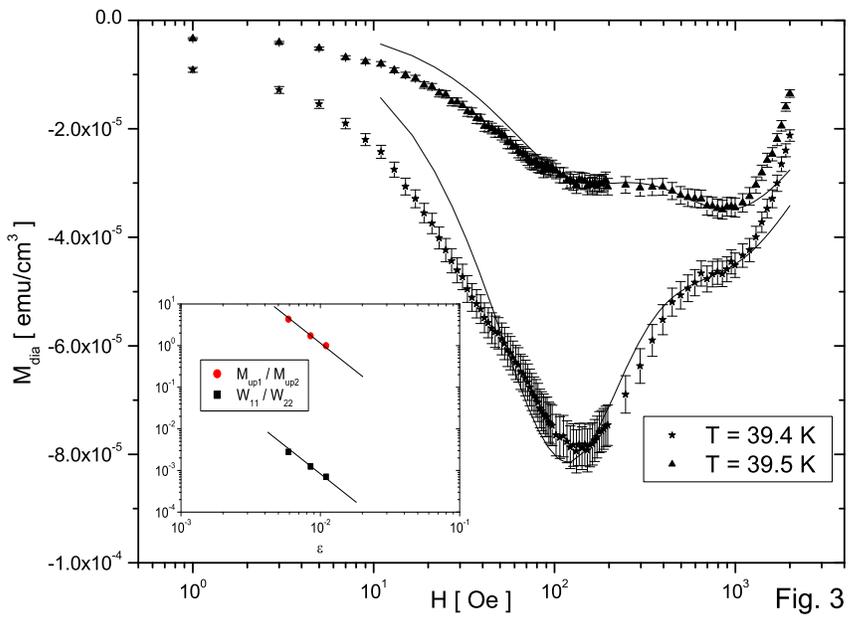
\end{document}